\documentclass[a4paper,fleqn,usenatbib]{mnras}

\usepackage{newtxtext,newtxmath}

\usepackage[T1]{fontenc}
\usepackage{ae,aecompl}

\usepackage{graphicx}	
\usepackage{amsmath}	
\usepackage{amssymb}	

\title[Tidal evolution of exoplanetary systems]{Tidal evolution of exoplanetary systems hosting Potentially Habitable Exoplanets. The cases of LHS-1140 b-c and K2-18 b-c}

\author[G. O. Gomes et al.]{
G. O. Gomes,$^{1,2}$\thanks{E-mail: gabrielogomes@usp.br}
S. Ferraz-Mello,$^{1}$
\\
$^{1}$Instituto de Astronomia, Geof\'isica e Ci\^encias Atmosf\'ericas, IAG-USP, Rua do Mat\~ao 1226, 05508-900 S\~ao Paulo, Brazil\\
$^{2}$Observatoire de Gen\`eve, Universit\'e de Gen\`eve, 51 Chemin des Maillettes, CH-1290 Sauverny, Switzerland
}

\date{Accepted 2020 April 19. Received 2020 April 19; in original form 2020 March 4}

\pubyear{2020}

\begin{document}
\label{firstpage}
\pagerange{\pageref{firstpage}--\pageref{lastpage}}
\maketitle

\begin{abstract}
We present a model to study secularly and tidally evolving three-body systems composed by two low-mass planets orbiting a star, in the case where the bodies rotation axes are always perpendicular to the orbital plane. The tidal theory allows us to study the spin and orbit evolution of both stiff Earth-like planets and predominantly gaseous Neptune-like planets. The model is applied to study two recently-discovered exoplanetary systems containing potentially habitable exoplanets (PHE): LHS-1140 b-c and K2-18 b-c. For the former system, we show that both LHS-1140 b and c must be in nearly-circular orbits. For K2-18 b-c, the combined analysis of orbital evolution timescales with the current eccentricity estimation of K2-18 b allows us to conclude that the inner planet (K2-18 c) must be a Neptune-like gaseous body. Only this would allow for the eccentricity of K2-18 b to be in the range of values estimated in recent works ($e=0.20 \pm 0.08$), provided that the uniform viscosity coefficient of K2-18 b is greater than $2.4 \times 10^{19} \ \textrm{Pa s}$ (which is a value characteristic of stiff bodies) and supposing that such system has an age of some Gyr. 
\end{abstract}




\section{Introduction}

After the discovery of the first exoplanet orbiting a solar-type star in 1995 \citep{Mayor1995}, a new research branch in astronomy related to the detection and characterization of exoplanetary systems ensued. Several projects have been developed with the aim of discovering exoplanets. On the one hand, missions based on the transit photometry technique (e.g Kepler, TESS, WASP) allow for a determination of the radii of exoplanets by using space or ground telescopes. On the other hand, high-precision spectrographs (e.g HARPS, HARPS-N, ESPRESSO, HIRES) are used to estimate the masses of exoplanets by the method of Doppler spectroscopy (a.k.a radial-velocity measurements). The combination of the data from Doppler spectroscopy measurements and transits thus allows for the estimation of exoplanets densities and bulk compositions. The diversity of masses and radii of the exoplanets discovered from these missions allows for the characterization of such bodies to vary between small Earth-like rocky planets or waterworlds to hot Jupiters and brown dwarfs (see e.g \citet{Kuchner2003,Leger2004} and references therein). Taking into account such diversity of the exoplanets compositions, several models have been developed with the objective of modelling their interior structure \citep{Seager2007,Adams2008,Batygin2013}. Tidal interactions and their resulting consequences for orbital evolution have been studied for both exoplanets \citep{Barnes2017,Barr2018} and their host stars \citep{Bolmont2012} in order to determine the fate of potentially habitable exoplanets (henceforth PHE) w.r.t their position in the habitable zone (henceforth HZ) of their host stars.

\citet{Barnes2017} has shown that the tidal locking is a major factor in the orbital evolution of PHE, and rotational synchronization may be a characteristic of the majority of these planets. The results presented by Barnes were based on simulations of the coupled spin-orbit tidal evolution of exoplanetary systems, where the Constant Time Lag (CTL) and Constant Phase Lag (CPL) Darwinian approaches were employed. 

In parallel works, \citet{Bolmont2017} and \citet{Gallet2017} studied the effects of stellar tidal dissipation on the evolution of close-in massive planets. Their results show that the dynamical tide (i.e the tidal interactions arising as a consequence of the excitation of inertia waves in the interior of the stars) may be the dominant effect on the orbital evolution of the exoplanetary systems in the Pre-Main Sequence (PMS), whereas the equilibrium tide (i.e the large-scale hydrodynamic adjustment of a body and the resulting flow as a consequence of the gravitational field of a companion) rules the orbital evolution when the star reaches the Zero Age Main Sequence (ZAMS) and evolves until the Red Giant Branch (RGB) phase. 

In most of the aforementioned works related to the study of tidal interactions and the subsequent orbital evolution of exoplanetary systems, Darwinian approaches have been employed to compute the effects of the equilibrium tides in both the exoplanets and their host stars. Such approaches require the knowledge of the tide lag, which is related to the quality factor ($Q$) ruling the energy dissipated by the bodies, and varies significantly among  predominantly rocky Earth-like planets and gaseous bodies such as hot Jupiters and host stars \citep{Dobbs-Dixon2004,Kellermann2018}. Moreover, the quality factor is an ad-hoc parameter that has not been rigorously linked to the internal structure and physical parameters of the bodies, as a consequence of the lack of a thorough knowledge of the dominant physical processes determining the magnitude of the dissipation factor responsible for the tidal evolution of planetary systems \citep{Dobbs-Dixon2004}. 

Recently, a new theory to model tidal interactions among celestial bodies was developed by \citet{Ferraz-Mello2013}. This theory (hereafter referred to as the creep tide theory) considers the deformations of an extended body due to the perturbation caused by the existence of a point mass companion. In this framework, the non-instantaneous response of the extended body's figure to the disturbing potential caused by the point mass and the rotation of the primary is ruled by the relaxation factor $\gamma$, which is inversely proportional to the uniform viscosity coefficient $\eta$ of the extended body. The resulting expressions for both the rotational and orbital evolution of the system depend on well-defined physical parameters. No ad-hoc constants ensue relating the tidal lags to the frequencies. Moreover, the phenomenon of capture in spin-orbit resonances may be described without the additional hypothesis of a permanent equatorial asymmetry.

In 2018, a new version of the creep tide theory was developed by Folonier et al. In this new version, the resulting equilibrium figure of the extended body is assumed to be an ellipsoid with unknown flattenings and orientation. The time evolution of the extended body's figure is obtained by the simultaneous integration of three first-order ordinary differential equations. The expression for the potential resulting from the deformations of the extended body is used to obtain the equations which rule the orbital evolution of the system and the rotational evolution of the extended body \citep{Folonier2018}, by employing the basic principles of Newtonian Mechanics. The old version of the creep tide theory \citep{Ferraz-Mello2013} predicted small-amplitude forced oscillations of the rotation in the case of stiff bodies in synchronous motions. The new version of the theory \citep{Folonier2018} takes them into account in a self-consistent way.

In this work, we describe a model to study the orbital and rotational evolution of three-body secularly evolving exoplanetary systems composed by a star and two low-mass planetary companions, such as super-Earths and mini-Neptunes. The model is only valid for the coplanar case, where all the bodies rotation axes are perpendicular to the orbital plane at each instant. Both tidal and secular planetary interactions are considered in the model. Tidal interactions are taken into account by employing the creep tide theory \citep{Ferraz-Mello2013,Folonier2018}, while the secular interactions between the planets are computed by using the models of \citet{Mardling2002} and \citet{Mardling2007}. The model is applied to two exoplanetary systems, namely K2-18 b-c and LHS-1140 b-c, where the choice of these systems for study is based on the possible existence of a PHE in each system (namely, K2-18 b and LHS-1140 b). In both cases, the mean motion ratio of the planets indicates that no mean motion resonances influence the dynamics of the system. For the specific case of the K2-18 b-c exoplanetary system, the lack of an estimation of the radius of K2-18 c indicates that both the cases of a rocky super-Earth and a gaseous mini-Neptune compositions must be considered to study the tidal evolution of this system.

The models describing tidal and secular interactions are presented in Sec.\,\ref{sec:model}, where only the main aspects of the theories, which are required to obtain the equations ruling the spin-orbit evolution of the system, are outlined. In Sec.\,\ref{sec:lhs}, we briefly introduce the discoveries related to the LHS-1140 b-c system and apply the theory to study the evolution of the system. The application to the K2-18 b-c system is performed in Sec.\,\ref{sec:k2}. The discussions and conclusion of the work are presented in Sec.\,\ref{sec:discussions}.

\section{Model description}
\label{sec:model}

In this section, we present the tidal and secular evolution models used to obtain the equations ruling the time evolution of the exoplanetary systems. The models employed in this work were already presented in other papers. Thus, we only present the main points of their theoretical formulation. For a more detailed description, the reader is referred to \citet{Folonier2018} for the tidal interactions model and \citet{Mardling2007} for the secular interactions model.

\subsection{Tidal interactions} 

We consider an extended body of mass $m$ (primary) and a point mass $M$ (companion) whose instantaneous distance to the primary is $r(t)$. We comment that, in actual applications of the model, both the star and the planets may play the role of the primary.

The primary is assumed to rotate with an angular velocity $\Omega$ pointing in the $z$ direction, perpendicular to the orbital plane. The creep tide theory assumes a first-order linear differential equation for the instantaneous surface figure $\zeta$ of the primary. This equation is an approximate solution of the Navier-Stokes equation in spherical coordinates, supposing a low-Reynolds-number flow. The equation reads \citep{Ferraz-Mello2013,Ferraz-Mello2019}
\begin{equation}
\dot{\zeta} = \gamma (\rho - \zeta) ,
\end{equation}
where $\gamma$ is the relaxation factor given by
\begin{equation}
\gamma = \frac{R g d}{2 \eta}  ,
\end{equation} 
with $R$, $g$, $d$ and $\eta$ being the mean radius, surface gravity, mean density and uniform viscosity coefficient of the primary and $\rho$ is the surface figure of the primary corresponding to the inviscid case (a.k.a static tide). Supposing that the resulting surface figure can be approximated by a triaxial ellipsoid rotated of an angle $\delta$ w.r.t the companion, we obtain a system of differential equations ruling the figure evolution of the primary, given by \citep{Folonier2018}
\begin{equation}
\dot{\delta} = \Omega  - \dot{\varphi} - \frac{\gamma \epsilon _{\rho}}{2 \mathcal{E}_{\rho}} \sin 2 \delta ,
\label{delta-dot}
\end{equation}
\begin{equation}
\dot{\mathcal{E}}_{\rho} = \gamma \left( \epsilon _{\rho}  \cos 2 \delta - \mathcal{E}_{\rho} \right) ,
\end{equation}
\begin{equation}
\dot{\mathcal{E}}_{z} = \gamma  \left( \epsilon _{z}  - \mathcal{E}_{z} \right) ,
\label{ez-dot}
\end{equation}
where $\mathcal{E}_{\rho}$ and $\mathcal{E}_{z}$ are the ellipsoid instantaneous equatorial and polar flattening coefficients, respectively, $\varphi$ is the true anomaly of the companion and $\epsilon_{\rho}$, $\epsilon _z$ are the flattenings in the inviscid case. They are given by
\begin{equation}
\epsilon _{\rho} = \frac{15}{4} \frac{M}{m} \frac{R^3}{r^3} \equiv \bar{\epsilon}_{\rho} \left( \frac{a}{r} \right) ^3,
\end{equation}
\begin{equation}
\epsilon _z = \frac{\epsilon _{\rho}}{2} + \bar{\epsilon}_z = \frac{\epsilon _{\rho}}{2} + \frac{5 \Omega ^2R^3}{4Gm} ,
\end{equation}
where $G$ is the gravitational constant. It is worth emphasizing that the assumption that the resulting equilibrium figure can be represented by a triaxial ellipsoid is an approximation. Such approximation is reasonable in the applications of this work since the planets are relatively far from their host star. However, such approximation may not be valid in the case of close-in planets subjected to large tidal distortions (see e.g the discussions presented in \citet{Hellard2019}).

The expression for the potential acting on the companion considering the resulting triaxial shape of the primary rotated of an angle $\varphi _{B} = \varphi + \delta $ w.r.t the axis x can be approximated, to first order in the flattenings, by
\begin{equation}
MU = - \frac{GMm}{r} - \frac{3GCM}{4r^3} \left[ \mathcal{E}_{\rho} \cos (2 \varphi _B - 2 \varphi ) + \frac{2}{3} \mathcal{E} _z \right] ,
\end{equation}
where $C$ is the moment of inertia of the primary.

Since we have the expression for the potential, we can calculate the components of the force acting on the companion ($\vec{F}$). Moreover, from the force expression, the torque can be obtained. The reaction to the torque acting on the companion is the torque ruling the rotational evolution of the primary, while the expressions for the derivative of the work done by the disturbing tidal potential ($\dot{W} = \vec{F} \cdot \vec{V}$, with $\vec{V}$ being the velocity of the companion) and the angular momentum variation ($\dot{\mathcal{L}}$) of the orbit give the equations ruling the semimajor axis and eccentricity evolution. The calculations are straightforward and we obtain
\begin{equation}
\dot{\Omega} = -\frac{ 3 G M }{2 r^3} \mathcal{E}_{\rho} \sin 2 \delta ,
\label{omegadot-complete}
\end{equation}
\begin{equation}
\dot{a} = \frac{2a^2}{GMm} \dot{W} ,
\label{dota}
\end{equation}
\begin{equation}
\dot{e} = \frac{1-e^2}{e} \left( \frac{\dot{a}}{2a} - \frac{\dot{\mathcal{L}}}{\mathcal{L}} \right) .
\label{dote}
\end{equation}

Eqs.\,\ref{omegadot-complete} -\,\ref{dote} must be solved simultaneously with Eqs.\,\ref{delta-dot} -\,\ref{ez-dot} to give the complete tidal evolution of the system. This method is henceforth referred to as full approach. We have six first-order ordinary differential equations to be solved. The number of differential equations to be solved can be reduced if we consider the constant rotation rate approximation. In this case, $\Omega$ is assumed to be a constant when solving for Eqs.\,\ref{delta-dot}-\ref{ez-dot}. Afterwards, the equation for $\dot{\Omega}$ is obtained from the torque expression. This approach is henceforth referred to as constant rotation rate approximation. In this case, the formulation is strictly equivalent to the previous version of the creep tide theory \citep{Ferraz-Mello2013,Ferraz-Mello2015a} and virtually equivalent to the approach of \citet{Correia2014}, which is based on a Maxwell viscoelastic rheology.

\subsubsection{Constant rotation rate approximation}
\label{sec:constantomega}

We now briefly mention some results of the constant rotation rate approximation.

Considering that the short-period variations of the rotation rate are negligible in Eq.\,\ref{delta-dot} (which is a reasonable assumption for a body far from the synchronous rotation regime), we can obtain a series expression for the differential equations ruling the orbital evolution of the system and the rotational evolution of the primary \citep{Gomes2019}. The calculations are straightforward and the results are the same as the ones presented in \citet{Ferraz-Mello2015a}. We have
	\begin{equation}
     \langle \dot{a} \rangle = \frac{R^2 n\bar{\epsilon}_{\rho}}{5a} \sum_{k \in \mathbb{Z}} \left[ 3 (2-k) \frac{\gamma (\nu+kn) E_{2,k}^2}{\gamma^2 + (\nu+kn)^2} - \frac{\gamma k^2 n E_{0,k}^2}{\gamma^2+k^2n^2}\right] ,
     \label{fm2015semi}
	\end{equation}
	\begin{equation}
	\langle \dot{e} \rangle = -\frac{3GMR^2\bar{\epsilon}_{\rho}}{10na^5e}\sum_{k \in \mathbb{Z}} \left[ P_k^{(1)} \frac{\gamma (\nu+kn) E_{2,k}^2}  {\gamma^2 + (\nu+kn)^2} + \frac{P_k^{(2)}}{3} \frac{\gamma k^2 n E_{0,k}^2}{\gamma^2 + k^2n^2} \right] ,
	\label{fm2015e}
	\end{equation}
	\begin{equation}
\langle \dot{\Omega} \rangle = - \frac{3GM\bar{\epsilon} _{\rho}}{2 a^3} \sum _{k \in \mathbb{Z}} \frac{\gamma (\nu + k n) E_{2,k} ^2}{{\gamma ^2 + (\nu + k n)^2 }} ,  
\label{average-omegadot}
\end{equation} 
	with $\nu = 2 \Omega - 2n$ being the semidiurnal frequency. Moreover, we have
	\begin{equation}
	P_k^{(1)}	=	\left[ 2 \sqrt{1-e^2} - (2-k)(1-e^2) \right] ,
	\end{equation}
	\begin{equation}
	P_k^{(2)}	=	1-e^2 ,
	\end{equation}
	and $E_{j,k}$ are Cayley coefficients, given by \citep{Ferraz-Mello2013,Ferraz-Mello2015a}
	\begin{equation}
	E_{2,k}(e) = \frac{1}{2\pi\sqrt{1-e^2}} \int_0^{2\pi} \frac{a}{r} \cos [2 \varphi + (k-2) \ell] d\varphi,
	\end{equation}
	where $\ell$ is the mean anomaly of the companion.
	
The above formulation adopting a constant rotation rate significantly reduces the time required to numerically integrate the equations of the orbital and rotational evolution of the system. The discrepancies between the constant rotation rate approximation and the full approach arise when the rotation is near the synchronous regime, for stiff bodies. An analytical approximation to correctly describe the figure and rotational evolution equations of the extended body in the synchronous rotation regime has already been developed by \citet{Folonier2018}, by taking into account the non-negligible short-period variations of the rotation rate. The results are briefly revisited in the next subsection. 

	\subsubsection{The synchronous regime}
	\label{sec:synchronous}
	
When the rotation of the extended body is damped to the synchronous (or pseudo-synchronous) attractor, the time evolution of the rotation and the ellipsoid's shape can be described by a sum of periodic components with frequencies $kn$, where $k \geq 1$ and $n$ is the orbital mean-motion \citep{Folonier2018}. The analytical expressions for $\Omega$, $\mathcal{E} _{\rho}$, $\mathcal{E}_{z}$ and $\delta$ allow for the determination of compact equations ruling the evolution of both the semimajor axis and the eccentricity. The resulting expressions are (to order $e^2$)

\begin{equation}
\langle \dot{a} \rangle = - \frac{21C \bar{\epsilon}_{\rho} e^2}{ma} \frac{n^2 \gamma}{n^2 + \gamma ^2} ,
\label{semi-sync}
\end{equation}
\begin{equation}
\langle \dot{e} \rangle = - \frac{21 C \bar{\epsilon}_{\rho} e (1-e^2)}{2ma^2} \frac{n^2 \gamma}{n^2 + \gamma ^2} .
\label{e-sync}
\end{equation}

\subsection{Secular interactions}
\label{sec:secular}

Additionally to the tidal interactions, the secular interactions between the planets must be taken into account in the case of a star with two planetary companions. For a coplanar system, the secular interactions cause a variation of the planets eccentricity as well as their longitudes of periastron $\varpi$. The model of \citet{Mardling2002} gives a system of four differential equations ruling the planets secular evolution. The equations are

\begin{equation}
\dot{e}_1 = -\frac{15}{16} n_1e_2 \frac{m_2}{m_0} \left( \frac{a_1}{a_2} \right) ^4 \frac{\sin \Delta \varpi}{(1-e_2^2)^{5/2}} ,
\label{secular-ec}
\end{equation}

\begin{equation}
\dot{e}_2 = \frac{15}{16} n_2e_1 \frac{m_1}{m_0} \left( \frac{a_1}{a_2} \right) ^3 \frac{\sin \Delta \varpi}{(1-e_2^2)^2} ,
\label{secular-eb}
\end{equation}

\begin{equation}
\dot{\varpi}_1 = \frac{3}{4}  \frac{n_1m_2}{m_0}  \left( \frac{a_1}{a_2} \right)^3 (1-e_2^2)^{3/2} \left[ 1 - \frac{5}{4} \left( \frac{a_1e_2}{a_2e_1} \right)  \frac{\cos \Delta \varpi}{1-e_2^2} \right] ,
\label{secular-varc}
\end{equation}

\begin{equation}
\dot{\varpi}_2 = \frac{3}{4}  \frac{n_2m_1}{m_0}  \left( \frac{a_1}{a_2} \right)^2 (1-e_2^2)^{-2} \left[ 1 - \frac{5}{4} \left( \frac{a_1e_1}{a_2e_2} \right)  \frac{1+4e_2^2}{1-e_2^2} \cos \Delta \varpi \right] ,
\label{secular-varb}
\end{equation}
where $\Delta \varpi = \varpi _1 - \varpi _2$, $m_0$ refers to the mass of the star and the parameters with subscript $i = 1$ ($2$) correspond to the inner (outer) planetary companion.

In practice, the equations for the secular evolution of the system are combined with the equations for the tidal evolution, thus giving a more complete description of the evolution of the system.
	
\section{LHS-1140 b-c. A system with two rocky super-Earths}
\label{sec:lhs}
	
LHS-1140 \textbf{b-c are} hosted by a mid-M dwarf. The current values for the estimated mean densities of the planets are $\rho _b = 7.5 \pm 1.0 \ \textrm{g cm}^{-3}$ \citep{Dittmann2017} and $\rho _c = 4.7 \pm 1.1 \ \textrm{g cm}^{-3}$ \citep{Ment2019}. The values of $\rho _b$ and $\rho _c$ suggest that both LHS-1140 b and LHS-1140 c may be rocky Earth-like planets. The hypothesis of a rocky structure was suggested for LHS-1140 b by \citet{Dittmann2017}., while the hypothesis of a rocky LHS-1140 c with a magnesium and silicate core was discussed by \citet{Ment2019}. Table\,\ref{table1} shows the values of some physical and orbital parameters taken from \citet{Ment2019}, which were used in this work to study the spin-orbit evolution of the system.

	\begin{table}
  \begin{center}
    
    \begin{tabular}{cc} 
      \hline
      \textbf{LHS-1140} & \textbf{Parameter value}  \\ 
      \hline
      
      Stellar mass ($M_{\odot}$) & $0.179 \pm 0.014$ \\
      Stellar radius ($R_{\odot}$) & $0.2139 \pm 0.0041$ \\
      Luminosity ($L_{\odot}$) & $0.00441 \pm 0.00013$ \\
      Effective temperature (K) & $3216 \pm 39$ \\
      Rotation period (days) & $131 \pm 5$ \\ 
      Age (Gyr)	&	$> 5$ \\
      \hline
      \textbf{LHS-1140 c} & \textbf{Parameter value}  \\ 
      \hline
      Planet mass ($M_{\oplus}$) & $1.81\pm0.39$ \\
	  Planet radius ($R_{\oplus}$) & $1.282\pm0.024$ \\
	  Mean density (g cm$^{-3}$) & $4.7 \pm 1.1$ \\ 
	  Semimajor axis (AU) & $0.02675\pm0.00070$ \\
	  Orbital period (days) & $3.777931\pm0.000003$ \\
	  Orbital eccentricity & $< 0.31 $ \\ \hline
      \textbf{LHS-1140 b} & \textbf{Parameter value}  \\ 
	  \hline
      Planet mass ($M_{\oplus}$) & $6.98\pm0.89$ \\
      Planet radius ($R_{\oplus}$) & $1.727\pm0.032$ \\
      Mean density (g cm$^{-3}$) & $7.5 \pm 1.0$ \\ 
	  Semimajor axis (AU) & $0.0936\pm0.0024$ \\
	  Orbital period (days) & $24.736959\pm0.000080$ \\
	  Orbital eccentricity & $< 0.06 $ \\ \hline
      
    \end{tabular}
    \caption{Parameters for the LHS-1140 system after \citet{Ment2019}. The eccentricity estimations have a $90\%$ confidence.} 
    \label{table1}
  \end{center}
\end{table}

\subsection{Spin-orbit resonances}

It is well-known that rocky bodies may be captured in spin-orbit resonances when the rotation evolves from initially fast rotating states \citep{Makarov2013,Correia2014,Ferraz-Mello2015a}. Tidal interactions slow down the body's rotation until an equilibrium configuration of the rotation is reached, where $\langle \dot{\Omega} \rangle = 0$, i.e the secular variation of the rotation is zero (the same mathematical condition holds for the secular torque, since it is proportional to $\dot{\Omega}$).

We now explore the possible spin-orbit resonances for LHS-1140 c and b by supposing that their current eccentricities are near the upper boundary values estimated by \citet{Ment2019}, namely $e_b = 0.06$ and $e_c = 0.31$ (cf. Table\,\ref{table1}). 

\begin{figure}
\centering
\includegraphics[height=100pt,width=\hsize]{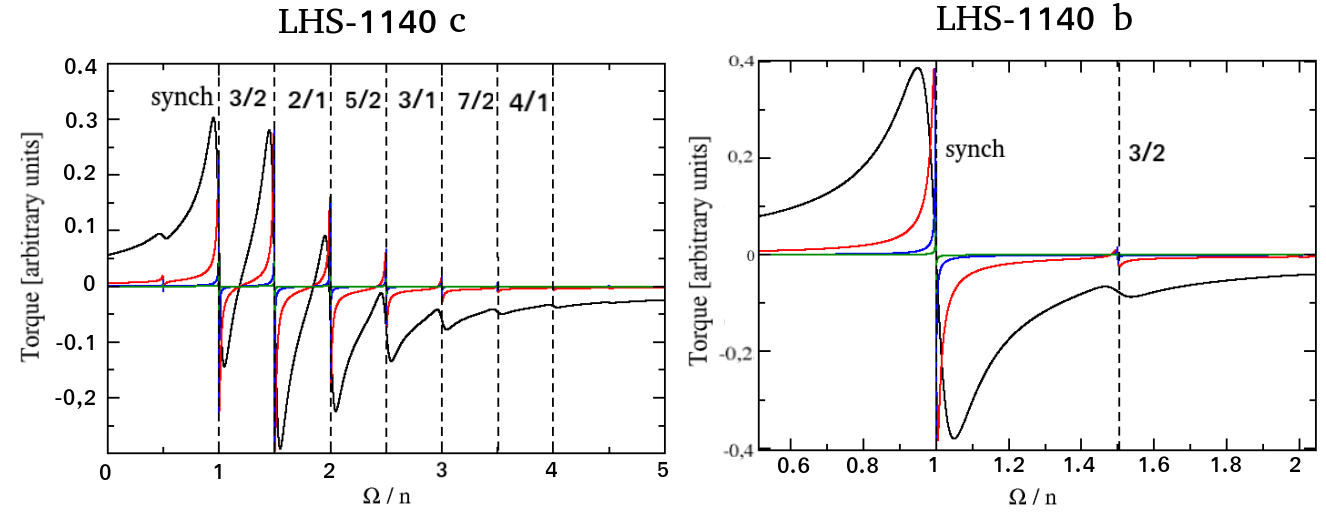}
\caption{Normalized secular torque as a function of the rotation rate normalized by the mean-motion value. The eccentricity values were fixed at the upper boundary values of Table\,\ref{table1}. The black, red, blue and green curves correspond to $\gamma _i / n _i = 10^{-1}$, $10^{-2}$, $10^{-3}$ and $10^{-4}$, respectively. The black dashed lines indicate the exact values of the spin-orbit resonance commensurabilibies.}
\label{figlhs}
\end{figure}

Fig.\,\ref{figlhs} shows the normalized torque as a function of the rotation rate, normalized by the mean-motion of the planets. We see that multiple equilibrium configurations of the rotation (corresponding to the points where the curves reach the x axis) ensue when the relaxation factor is smaller (compare the different curves on each panel). The stable (unstable) equilibrium configurations of the rotation are the descending (ascending) intersections of the normalized torque function with the horizontal axis. For LHS-1140 c, considering $\gamma / n = 0.1$ (corresponding to $\gamma _c = 1.93 \times 10^{-6} \ \textrm{s}^{-1}$), we see that there are three stable resonant spin-orbit configurations, with $\Omega / n = 1$, $\Omega / n = 3/2$ and $2/1$ (cf. black curve on the left panel of Fig.\,\ref{figlhs}). The number of stable spin-orbit resonances increases as $\gamma / n$ decreases. For LHS-1140 b, the number of stable spin-orbit resonances is smaller due to the smaller eccentricity of the planet, and even for $\gamma / n = 10^{-4}$ (corresponding to $\gamma _b = 2.95 \times 10^{-10} \ \textrm{s}^{-1}$), only the 3/2 and the synchronous spin-orbit resonances are stable. It is worth mentioning that, in Fig.\,\ref{figlhs}, the time evolution of the rotation rate is represented by the decrease (increase) of the rotation rate in the case of the initially fast-rotating prograde (retrograde) case.

The results of Fig.\,\ref{figlhs} show that, if the eccentricities of LHS-1140 b and c are close to the upper boundary values estimated by \citet{Ment2019}, the rotation of the planets is most likely not synchronous if we suppose that the planets had a past fast rotation rate with $\Omega \gg n$.

\subsection{Spin-orbit evolution}

We now study the tidal evolution of the planets orbits to assess the timescales of evolution and the effect of the coupled tidal and secular interactions between the planets. Since the current rotational configuration of the exoplanets is not known, we consider both the initially fast-rotating and synchronous cases to show a more complete description of the possible spin-orbit evolution scenarios of the planets. 

Several numerical integrations of the differential equations ruling the secular and tidal evolution of the system were performed. For the sake of making clear the contributions of the secular and tidal interactions on the time evolution of the system, we compared the results of the secular model to simulations where only the tidal interactions were considered. In all the calculations performed in this section, we neglect the effects of the stellar tides (i.e the tides raised in the star by the planets), since some numerical experiments have shown that the stellar tide effects are negligible compared to the planetary tide effects, due to the planetary companions low masses (such results have also been obtained in other works, see e.g \citet{Rodriguez2011}).

\begin{figure}
\centering
\includegraphics[height=200pt,width=\hsize]{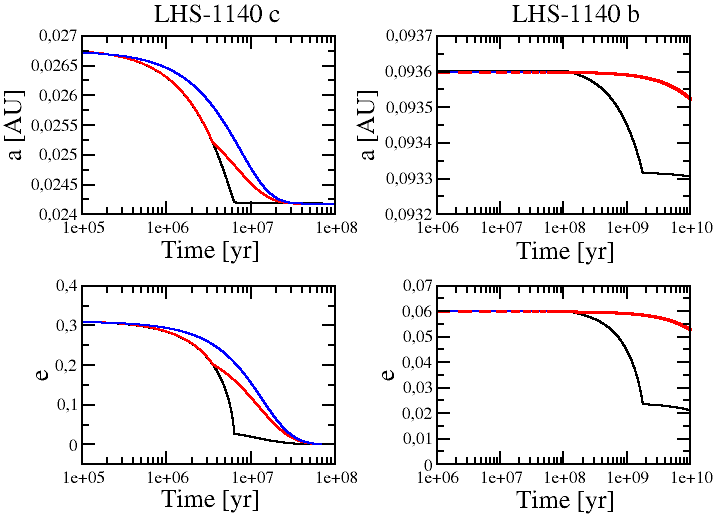}
\caption{Semimajor axis and eccentricity evolutions of both LHS-1140 c and b, for $\gamma _c = 10^{-7} \ \textrm{s}^{-1}$ and $\gamma _b = 10^{-8} \ \textrm{s}^{-1}$. In these panels, we have (i) the initially fast prograde rotation case (black), (ii) the initially fast retrograde rotation case (red) and (iii) the initially synchronous case (blue). For LHS-1140 b, the initially synchronous case gives the same results as the initially fast-rotating retrograde case.}
\label{fig8}
\end{figure}

\begin{figure}
\centering
\includegraphics[height=100pt,width=\hsize]{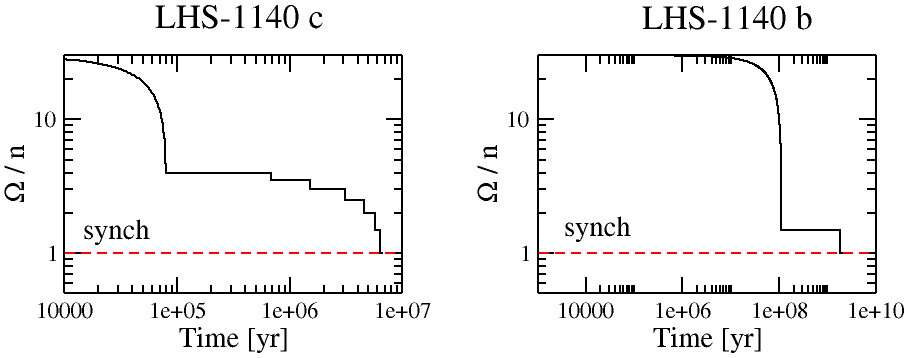}
\caption{Rotational evolution of LHS-1140 c and b, corresponding to the initially fast-rotating prograde cases of Fig.\,\ref{fig8}. The initial rotation was taken such that $\Omega / n = 30$. The red dashed lines correspond to the synchronous rotation value $\Omega = n$.}
\label{fig:rotlhs}
\end{figure}

Fig.\,\ref{fig8} shows a scenario of the tidal evolution of LHS-1140 b and c, considering no secular interactions between the planets. We used $\gamma _b = 10^{-8} \ \textrm{s}^{-1}$ and $\gamma _c = 10^{-7} \ \textrm{s}^{-1}$ (which are values close to the characteristic values used for rocky planets, see \citet{Ferraz-Mello2013} Table 1), where a larger relaxation factor value was attributed to LHS-1140 c given its smaller mean density value when compared to LHS-1140 b. Three initial values of the rotation rate were considered for LHS-1140 c, given by: (i) the initially fast-rotating prograde case (black curve); (ii) the initially fast-rotating retrograde case (red curve); and (iii) the initially synchronous case (blue curve). For LHS-1140 b (right panels), we have: (i) the initially fast-rotating prograde case (black curve); and (ii) the initially fast-rotating retrograde case (red curve). In this case, the initially synchronous case gives the same results as the initially fast-rotating retrograde case, since no other spin-orbit resonances are possible other than synchronism, due to the smaller initial eccentricity value for this planet (see Fig.\,\ref{figlhs}). By analysing the results of Fig.\,\ref{fig8}, we see that the eccentricity damping occurs more rapidly in the fast rotating cases (black curves), when compared to the initially synchronous configuration. The capture in the synchronous regime can be detected by analysing the eccentricity decay curve. Indeed, in all panels of Fig.\,\ref{fig8}, there is a characteristic elbow in all the curves corresponding to the initially fast rotating cases. This elbow corresponds to the point where the planets rotation reaches synchronism. Such visible signature of the capture in the synchronous rotation regime was already discussed by other authors \citep{Rodriguez2012}. The spin evolution of the planets in the initially fast-rotating prograde case of Fig.\,\ref{fig8} can be seen in Fig.\,\ref{fig:rotlhs}. As it is expected, the time of capture in the synchronous regime (cf. Fig.\,\ref{fig:rotlhs}) corresponds to the time where the elbow is seen in the panels of Fig.\,\ref{fig8}.

We also emphasize the staircase behavior of the rotation rate as a function of time in Fig.\,\ref{fig:rotlhs}. This behavior is a consequence of the fact that the graph of the torques vs. the rotation rate shows a succession of kinks that act as barriers for the evolution of the rotation. For instance, consider the case shown by the black curve in Fig.\,\ref{figlhs} (left). One body whose rotation velocity is initially high (i.e $\Omega \gg n$) will evolve leftwards (the torque is negative) until it reaches the $2/1$ spin-orbit resonance, where the torque sign changes and the rotation can no longer evolve. The body remains trapped in that resonance. However, when the eccentricity decreases, the height of the peak at that kink also decreases. As the tidal evolution of the eccentricity is continuously decreasing, the height of the peak may become negative and the barrier of positive torques disappear. The rotation will again evolve leftwards up to reach the next resonance (in the considered example, the $3/2$ spin-orbit resonance). This behavior will repeat itself up to the point where the rotation is trapped into the synchronous rotation rate state. Such kinks of the torques are also responsible for speeding up the process of rotation rate decay when the rotation approaches the neighborhood of spin-orbit resonances (the torques are higher when $\Omega$ approaches spin-orbit resonances). Such evolutionary behavior is characteristic of stiff bodies (where $\gamma \ll n$).

Several other numerical integrations to study the tidal evolution of the planets, considering no secular interactions between them, were performed, and we varied the planets relaxation factor values. We verified that the timescale for which orbital circularization takes place scales with $\gamma^{-1}$ for the characteristic range of values of the relaxation factor of stiff bodies, namely $\gamma \ll n$.

\begin{figure}
\centering
\includegraphics[height=200pt,width=\hsize]{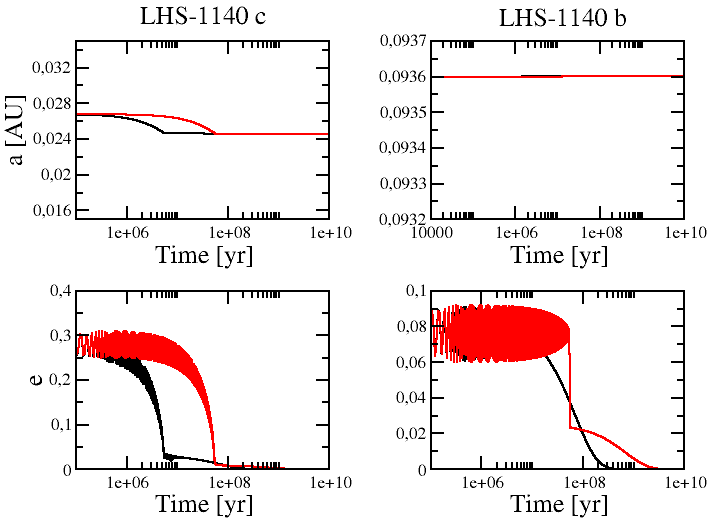}
\caption{Orbital evolution of both LHS-1140 c and b, considering secular and tidal interactions. The black curves show an evolution scenario with $\gamma _c = 10^{-7} \ \textrm{s}^{-1}$ and $\gamma _b = 10^{-8} \ \textrm{s}^{-1}$, while the red curves show a scenario when $\gamma _c = 10^{-8} \ \textrm{s}^{-1}$ and $\gamma _b = 10^{-9} \ \textrm{s}^{-1}$.}
\label{fig:orblhs-sec}
\end{figure}

Fig.\,\ref{fig:orblhs-sec} shows two scenarios of the orbital evolution of LHS-1140 c and b considering both tidal and secular interactions between the planets. We have $\gamma _c = 10^{-7} \ \textrm{s}^{-1}$ and $\gamma _b = 10^{-8} \ \textrm{s}^{-1}$ on the black curves, and $\gamma _c = 10^{-8} \ \textrm{s}^{-1}$ and $\gamma _b = 10^{-9} \ \textrm{s}^{-1}$ on the red curves. The characteristic decay of the eccentricities due to the tidal interactions is present. The secular interactions are responsible for the oscillatory behavior of the planetary eccentricities. Additionally, the existence of the secular interactions cause an entanglement of the planetary eccentricities, and the characteristic timescale of orbital circularization of the inner planet (LHS-1140 c) rules the eccentricity decay of the outer planet (LHS-1140 b). This effect can be seen by comparing the timescales of orbital circularization of LHS-1140 b in Figs.\,\ref{fig8} and \ref{fig:orblhs-sec}.

One last aspect regarding the orbital evolution of the planets in the case of Fig.\,\ref{fig:orblhs-sec} is the lack of planetary migration for the outer planet. This effect is due to the decrease in the timescale of orbital circularization of the outer planet as a consequence of the coupling of the planetary eccentricities. The decay of the rotation rate to stable spin-orbit resonant states and the orbital circularization processes are no longer events that happen sequentially (which was the case in the purely tidally-evolving scenario). In the secularly-evolving case, these events take place on approximately the same timescale for the outer planet, and the most significant effect of the planetary migration (which happens only when the planet has already reached synchronism or non-synchronous spin-orbit resonances) does not take place.   
\par The relaxation factor of the planets was varied between $10^{-6} \ \textrm{s}^{-1}$ and $10^{-9} \ \textrm{s}^{-1}$, by maintaining the same ratio between the relaxation factor values of $\gamma _b / \gamma _c = 0.1$. We verified that the timescale for orbital circularization of the planets is inversely proportional to $\gamma _c$ when we consider both tidal and secular interactions. The existence of secular interactions between the planets causes an entanglement in the planetary eccentricities, which decreases the timescale of orbital circularization of the outer planet when compared to the case where no secular interactions are considered between the planets. We also verified that the timescale of orbital circularization is approximately 1 order of magnitude larger in the initially synchronous and fast-rotating retrograde cases, when compared to the initially fast-rotating prograde case.

We analysed the spin and orbit evolution of the system by considering that the two planets have a small value of the relaxation factor (which is consistent with the discussions presented in \citet{Dittmann2017} and \citet{Ment2019} regarding the planets compositions and internal structure models). The results of the numerical experiments have shown that the eccentricity decay of both planets is ruled by the eccentricity decay of the inner planet. The secular interactions between the planets cause a coupling between the planetary eccentricities, thus forcing the eccentricity decay of the outer planet at a time much smaller than the timescale of eccentricity decay due to the outer planet tidal interactions alone. However, the rotation rate and the semimajor axis evolution are not affected by secular interactions. Thus, the rotation of the outer planet may reach synchronization much later than the inner planet. In this case, if the synchronization of the outer planet is reached after orbital circularization has already taken place, the semimajor axis evolution of the outer planet may be neglected.

\section{K2-18 b-c. A system with a low-density super-Earth and a small inner planet}
\label{sec:k2}

K2-18 b was firstly detected in 2015 \citep{Montet2015}. In 2017, Cloutier et al. provided a constraint for the planet's density through combined data obtained from the HARPS and CARMENES instruments. In the latter work, the authors estimated a mean density of $\rho _b = 3.3 \pm 1.2 \ \textrm{g cm}^{-3}$ for this planet. Moreover, the existence of another planetary companion (namely, K2-18 c) in this system was firstly discussed. The existence of K2-18 c was confirmed in 2019, by Cloutier et al. In this same work, the physical and orbital parameters of K2-18 b were updated. The parameters used in the present work are given in Table\,\ref{table2}. All data were taken from \citet{Cloutier2019}, except from K2-18 b eccentricity value, which was taken from \citet{Sarkis2018}, and the orbital periods values of K2-18 b and c, which were taken from \citep{Montet2015} and \citet{Cloutier2017}, respectively.

	\begin{table}
  \begin{center}
    
    \begin{tabular}{cc} 
      \hline
      \textbf{K2-18} & \textbf{Parameter value}  \\ 
      \hline
      
      Stellar mass ($M_{\odot}$) & $0.495 \pm 0.004$ \\
      Stellar radius ($R_{\odot}$) & $0.469 \pm 0.010$ \\
      Effective temperature (K) & $3503 \pm 60$ \\
      Rotation period (days) & $39.63 \pm 0.50$ \\ 
      Age (Gyr)	&	$--$ \\
      \hline
	  \textbf{K2-18 c} & \textbf{Parameter value}  \\ 
	  \hline
	  Planet mass ($M_{\oplus} \times \sin i$) & $5.62\pm0.84$ \\
	  Planet radius ($R_{\oplus}$) & $--$ \\
	  Mean density (g cm$^{-3}$) & $--$ \\ 
	  Semimajor axis (AU) & $0.0670\pm0.0002$ \\
  	  Orbital period (days) & $8.962\pm0.008$ \\
	  Orbital eccentricity & $< 0.2 $ \\
      \hline
      \textbf{K2-18 b} & \textbf{Parameter value}  \\ 
      \hline
      Planet mass ($M_{\oplus}$) & $8.63\pm1.35$ \\
      Planet radius ($R_{\oplus}$) & $2.711\pm0.065$ \\
      Mean density (g cm$^{-3}$) & $2.4 \pm 0.4$ \\ 
	  Semimajor axis (AU) & $0.1591\pm0.0004$ \\
  	  Orbital period (days) & $32.94488\pm0.00281$ \\
	  Orbital eccentricity (*) & $0.20 \pm 0.08$ \\ \hline
      
    \end{tabular}
    \caption{Parameters for the K2-18 system after \citet{Cloutier2019}, except for K2-18 b eccentricity value, which was taken from \citet{Sarkis2018}, and the periods of the planets b and c, which were taken from \citep{Montet2015} and \citet{Cloutier2017}, respectively.} 
    \label{table2}
  \end{center}
\end{table}

The mean density of K2-18 b is currently estimated to be $\rho _b = 2.4 \pm 0.4 \ \textrm{g cm}^{-3}$ \citep{Cloutier2019}. Comparing such value with the estimated mean densities of the Earth ($\rho _{ear}$) and Neptune ($\rho _{nep}$), we see that $\rho _b$ is closer to $\rho _{nep}$, thus suggesting that a higher value of the relaxation factor (typical of gaseous bodies) would be the most reasonable assumption for such planet. However, taking into account the classification of K2-18 b as a PHE, we might expect an Earth-like composition for such planet, in which case we would have a small value for the relaxation factor with $\gamma _b$ in the interval $10^{-7} - 10^{-9} \ \textrm{s}^{-1}$.

For K2-18 c, an estimation of the planet mean radius is not currently available. Thus, we cannot estimate precisely the range of values of the relaxation factor of this planet. Since it was verified, in the previous section, that the eccentricity decay of secularly-evolving two-planets systems is ruled by the decay of the inner planet eccentricity, we analyse the orbital evolution considering both a Neptune-like gaseous structure and an Earth-like rocky structure for K2-18 c in order to obtain a more complete scenario of the timescales of orbital circularization for such system. 

\subsection{Case 1. A rocky K2-18 c}

In the case of a rocky K2-18 c with an Earth-like structure, we would have $R_c = 1.77 \ R_{\oplus}$ (which gives a mean density equal to the Earth's mean density value), and a range of values of $\gamma _c = 10^{-7} - 10^{-9} \ \textrm{s}^{-1}$ for the relaxation factor. Considering such range of values of $\gamma _c$, the rotation may be captured in non-synchronous spin-orbit resonances, provided the initial rotation rate of the body is such that $\Omega \gg n$. We considered several numerical integrations of the secular model for the evolution of the system, where the relaxation factor of K2-18 b was varied from $\gamma _b = 10^{-7} \ \textrm{s}^{-1}$ to $\gamma _b = 10^{-8} \ \textrm{s}^{-1}$. In all the cases, an initially fast-rotating state for both the planets was assumed with $\Omega _0 = 30 n$, and the eccentricities were set to $e_0 = 0.2$ \citep{Sarkis2018,Cloutier2019}.

\begin{figure}
\centering
\includegraphics[height=200pt,width=\hsize]{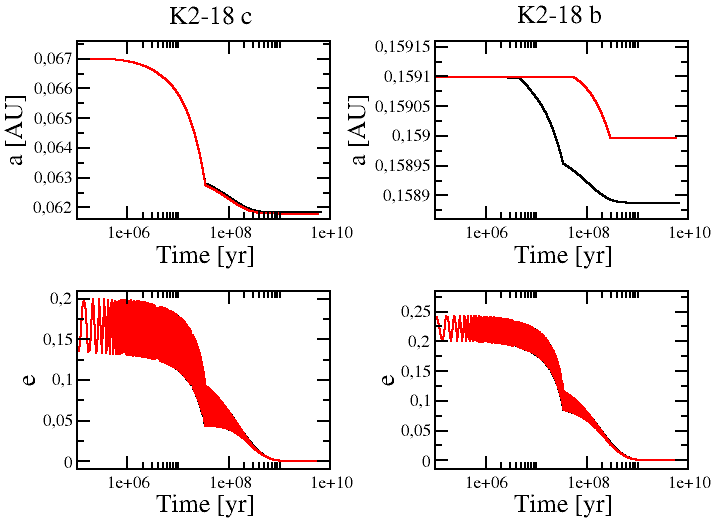}
\caption{Semimajor axis and eccentricity evolution of K2-18 \textbf{b-c} in the case of a rocky K2-18 c with a relaxation factor of $\gamma _c = 10^{-7} \ \textrm{s}^{-1}$. Two cases with different values for $\gamma _b$ are shown, and we have $\gamma _ b = 10^{-7} \ \textrm{s}^{-1}$ in the black curve and $\gamma _ b = 10^{-8} \ \textrm{s}^{-1}$ in the red curve.}
\label{fig:fig1-k218bc}
\end{figure}

Fig.\,\ref{fig:fig1-k218bc} shows the results of two numerical integrations of the tidal secular model. On the black curves, we set $\gamma _b = 10^{-7} \ \textrm{s}^{-1}$  and on the red curves we set $\gamma _b = 10^{-8} \ \textrm{s}^{-1}$. In both cases, we used $\gamma _c = 10^{-7} \ \textrm{s}^{-1}$. Both the rotation of K2-18 b and c are trapped in non-synchronous spin-orbit resonances since the initial values of the eccentricities of both planets are of the order $e=0.2$. As the eccentricity decay takes place, non-synchronous spin-orbit resonances no longer exist and both planets reach synchronism. The point where the rotation rate of the planets reaches synchronism corresponds to the point where the characteristic elbow is seen in all the panels of Fig.\,\ref{fig:fig1-k218bc} (such behavior was already discussed in the previous section). 

The timescale of orbital circularization of the system is independent of the relaxation factor of K2-18 b. In both cases shown in Fig.\,\ref{fig:fig1-k218bc}, the semimajor axis variation of the outer planet is relatively small.

\subsection{Case 2. A gaseous K2-18 c}

In the case of a predominantly gaseous structure for K2-18 c, the relaxation factor may be in the range $1 - 100 \ \textrm{s}^{-1}$. We chose the mean radius value of K2-18 c giving approximately the same mean density value of Neptune, corresponding to $R_c = 2.68 \ R_{\oplus}$.

\begin{figure}
\centering
\includegraphics[height=200pt,width=\hsize]{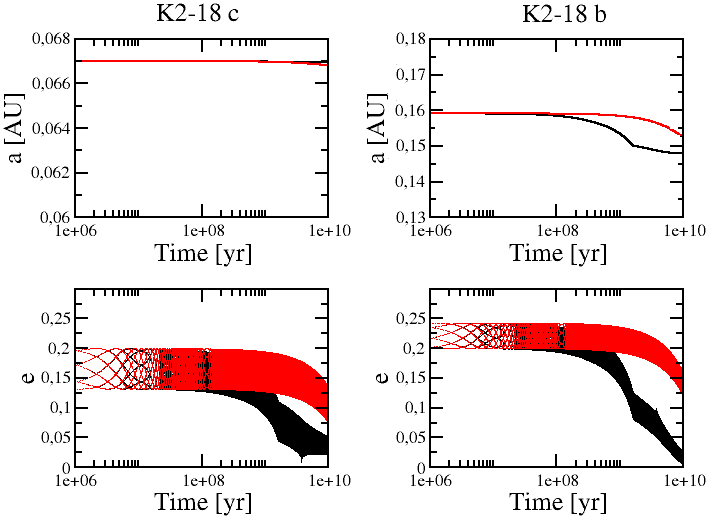}
\caption{Semimajor axis and eccentricity evolution of K2-18 \textbf{b-c} in the case of a gaseous Neptune-like K2-18 c with a relaxation factor of $\gamma _c = 1 \ \textrm{s}^{-1}$. Two cases with different values for $\gamma _b$ are shown, and we have $\gamma _ b = 10^{-7} \ \textrm{s}^{-1}$ in the black curve and $\gamma _ b = 10^{-8} \ \textrm{s}^{-1}$ in the red curve.}
\label{fig:fig2K218bc}
\end{figure}

Fig.\,\ref{fig:fig2K218bc} shows a scenario of the orbital evolution of K2-18 b-c considering $\gamma _b = 10^{-7} \ \textrm{s}^{-1}$ (black curve) and $\gamma _b = 10^{-8} \ \textrm{s}^{-1}$ (red curve). In both cases we used $\gamma _c = 1 \ \textrm{s}^{-1}$. In this case, we see that the orbital circularization timescale of the planets is no longer ruled by the inner planet eccentricity decay, but by the eccentricity decay of the outer planet. Such characteristic is a consequence of the fact that the eccentricity decay of the outer planet due to tidal interactions occurs more rapidly due to its relaxation factor value, despite the fact that its distance to the star is much bigger than the distance between the inner planet and the star. Since the eccentricities of the planets are coupled as a consequence of the secular interactions, the eccentricity decay of the outer planet forces the eccentricity decay of the inner planet.

Another interesting characteristic of the results shown in Fig.\,\ref{fig:fig2K218bc} is that a moderately high eccentricity value for the planets can be maintained for a time interval of some Gyr, provided that $\gamma _b = 10^{-8} \ \textrm{s}^{-1}$. Such scenario would not be possible if we considered a small value for the relaxation factor of K2-18 c (see Fig.\,\ref{fig:fig1-k218bc} and the discussion of Case 1). Thus, the comparison of the eccentricity decay timescales of Cases 1 and 2 taking into account the eccentricity estimation given by \citet{Sarkis2018} for K2-18 b (namely $e_b = 0.20 \pm 0.08$) allows us to conclude that the inner planet (K2-18 c) of the system may be predominantly gaseous, with a big value for the relaxation factor. Otherwise, orbital circularization would already have taken place on a timescale of approximately $0.1 \ \textrm{Gyr}$ (cf. Fig.\,\ref{fig:fig1-k218bc}).

\subsection{Discussion}

The lack of information on the radius of the inner planet forced us to study the orbital evolution of the system considering both a predominantly rocky (Case 1) and a predominantly gaseous (Case 2) structure for the inner planet. In both cases, we supposed a small value for the relaxation factor of K2-18 b due to the possibility of this planet being a PHE. In Case 1, we verified that the eccentricity damping of both planets is dominated by the inner planet's eccentricity decay (as it was the case of the LHS-1140 b-c system). The results show that orbital circularization is reached after approximately $0.1 \ \textrm{Gyr}$, supposing that the initial eccentricities of the planets are $e_0 = 0.2$ and $\gamma _c = 10^{-7} \ \textrm{s}^{-1}$. Moreover, the timescale of orbital circularization is independent of the value of $\gamma _b$. Analysing the eccentricity constraint for K2-18 b given by \citet{Sarkis2018} and the timescales of orbital circularization considering a rocky K2-18 c, we see that it is highly unlikely for K2-18 b to have an eccentricity of the order $0.1$ if the inner planet (K2-18 c) has a rocky composition. Such results show that the inner planet of the system is more likely a gaseous mini-Neptune with a large value of the relaxation factor (of the order $1 - 100 \ \textrm{s}^{-1}$), which is the scenario shown in Case 2. In such case, we verified that an eccentricity of $e_b = 0.2$ can be maintained for a timescale of some Gyr if $\gamma _b \leq 10^{-8} \ \textrm{s}^{-1}$.

We also performed other simulations considering an initially small eccentricity value for K2-18 c ($e_c \approx 0.05$) and a moderately high eccentricity for K2-18 b ($e_b \approx 0.2$) to investigate the possibility of having a stiff Earth-like nearly circular K2-18 c and a currently eccentric K2-18 b. However, the results have shown that the moderately high value of the eccentricity of K2-18 b would excitate the eccentricity of K2-18 c, thus leading to the same scenario shown in Case 1.

\section{Conclusion}
\label{sec:discussions}

We have used the creep tide theory to present a model describing the spin and orbit evolution of exoplanetary systems with two planetary companions whose internal structure composition may be described by two distinct regimes of the relaxation factor: (i) for predominantly gaseous planets, the relaxation factor was supposed to lie in the range of $1 - 100 \ \textrm{s}^{-1}$. (ii) For predominantly rocky planets, we supposed that the relaxation factor was varied between $10^{-7} - 10^{-9} \ \textrm{s}^{-1}$. Additionally to the planetary tidal interactions, we simultaneously computed the effects of secular interactions between the planets, in order to have a more realistic description of the mechanisms ruling the orbital evolution of the system. Stellar tide effects were neglected due to the small planetary masses (which lie in the interval of $1 - 10 \ \textrm{M}_{\oplus}$).

The secular model was applied to the LHS-1140 b-c and K2-18 b-c exoplanetary systems. For the LHS-1140 b-c exoplanetary system, the results have shown that, if both exoplanets are rocky planets (which is consistent with the discussions presented by \citet{Dittmann2017}), the timescales of rotational synchronization and eccentricity decay due to tidal interactions are much smaller than the estimated age of the system. Thus, we conclude that the planets are probably in nearly-circular orbits and with their rotation periods synchronized with their orbital periods. We emphasize that the orbital circularization of the planets would already have taken place even if we suppose higher values for the initial eccentricities used in the simulations, provided that the inner planet is considered as a rocky planet. Nevertheless, since no precise estimations of the current eccentricities exist, we cannot exclude the possibility of the planets being in moderately eccentric orbits. In that case, the planets would rather be characterized as Neptune-like gaseous planets. Thus, better constraints on the planets uniform viscosity coefficient (and consequently, the planets response to tidal stress) may be obtained when more precise estimations of the planets eccentricities exist. For the K2-18 b-c exoplanetary system, we compared the timescales of orbital and rotational evolution of the planets to the eccentricity estimations of the planets. The combination of these data has lead us to conclude that the inner planet cannot be a rocky super-Earth if we consider the eccentricity estimations for K2-18 b after \citet{Sarkis2018}, namely $e_b = 0.20 \pm 0.08$. The outer planet may be a rocky super-Earth with a relaxation factor of the order $\gamma _b \leq 10^{-8} \ \textrm{s}^{-1}$, in which case the orbital circularization would take place on a timescale of some Gyr. In all the cases analysed in this work, we have verified that the secular interactions couple the planets eccentricities, and the planet with the faster eccentricity decay rules the eccentricity decay of both planets. We emphasize that the initial values of the eccentricities of the planets were chosen to be the current estimated values. However, the results would hold for any other arbitrarily chosen initial eccentricity values.

Additionally to the analyses presented in this work for K2-18 b-c and LHS-1140 b-c, we also calculated the timescales of orbital and rotational evolution of some single-planet systems where the planet is currently classified as a PHE. The analyses were performed for Kepler-61 b, Kepler-440 b, Kepler-442 b and Kepler-443 b (the orbital and physical parameters of these systems were taken from \citet{Ballard2013} and \citet{Torres2015}). In all these cases, the lack of the mass estimation of the planets forced us to consider both the cases of gaseous and rocky bodies to study the tidal evolution of the planets. We observed that, due to the large values of the orbital periods of the planets (with $P_{orb} > 100 \ \textrm{days}$ for Kepler-440, 442 and 443 b and $P_{orb} \approx 59.8 \ \textrm{days}$ for Kepler-61 b), the timescales of orbital evolution are much larger than $1$ Gyr, independently of the value of the relaxation factor of the planets. Thus, no constraints on the relaxation factor of the planets can be obtained by analysing the eccentricity decay of these planets.

\section*{Acknowledgements}

This investigation is funded by the National Research Council, CNPq, Grant 302742/2015-8 and by FAPESP, Grants 2016/20189-9, 2017/25224-0 and 2019/21201-0. This investigation is part of the thematic project FAPESP 2016/13750-6.

\section*{Additional Information}

Competing Interests: The authors declare no competing interests.






\bsp	
\label{lastpage}
\end{document}